\documentclass[11pt,a4paper]{article}
\usepackage{jheppub}

\usepackage{physics}
\usepackage{tensor}
\usepackage{amsmath}
\usepackage{amssymb}
\usepackage[colorlinks=true]{hyperref}
\usepackage{xcolor}
\usepackage{cancel}
\usepackage{ytableau}

\usepackage{titlesec}
\makeatletter
\g@addto@macro\bfseries{\boldmath}
\makeatother

\ytableausetup{boxsize=0.3em,centertableaux}

\usepackage{framed}
\usepackage[most]{tcolorbox}
\usepackage{xcolor}
\colorlet{shadecolor}{gray!10}
\usepackage{marginnote} 

\newcommand{\re}[1] {(\ref{#1})}

\newcommand{\nn}{\nonumber}
\newcommand{\al}{\alpha} 
\newcommand{\be}{\beta}

\newcommand{\de}{\delta}

\newcommand{\beq}{\begin{equation}}
\newcommand{\eeq}{\end{equation}}

\preprint{UUITP-27/26} 

\title{$eAdS$ – embedded $AdS$,  tensionless strings and  contracted Carroll-Weyl symmetry
}

\author[a]{Ulf Lindstr\"{o}m}
\author[b]{Bo Sundborg}

\affiliation[a]{Department of Physics and Astronomy, Uppsala University,
Box 516, SE-75120 Uppsala, Sweden
and Centre for Geometry and Physics, Uppsala University,
Box 480, SE-75106 Uppsala, Sweden
}

\emailAdd{ ulf.lindstrom@physics.uu.se}

\affiliation[b]{The Oscar Klein Centre \& Department of Physics, Stockholm University,\\
AlbaNova, 106 91 Stockholm, Sweden}
\emailAdd{bo.sundborg@fysik.su.se}

\abstract{
We adapt the Dirac conformal $d+2$ dimensional embedding construction to define a null string in $AdS_{d+1}$. Introducing an extra St\''uckelberg field $\varphi$, which is a scalar under $AdS$ and shifts under local scaling of embedding space, the model acquires a local scaling invariance. We refer to this string as the embedded anti-De Sitter, $eAdS$, null string. A very useful property of this new gauged formulation is that, unlike previous gauged actions in $AdS$ background, the local scaling symmetry commutes with the $so(d,2)$ symmetry of $AdS$. Interestingly,  the $eAdS$ null string is the tensionless limit of a tensile string in embedding space and the same holds for its reduction to an intrinsic description of AdS.  We display the symmetries and constraint algebras both in  embedding space and directly in AdS.
}

\begin{document}
\pagestyle{myplain}
\maketitle

\section{Introduction}
 Strings with a null world sheet were introduced a long time ago, from two slightly different points of view \cite{Schild:1976vq,Karlhede:1986wb}. It was given a more geometric guise in \cite{Isberg:1993av}, lately dubbed the ILST string. Recently, the upsurge in interest in Carrollian geometry and Carroll-Weyl symmetry has led to a renewed interest in various aspects of this model, in particular in gauging the geometric formulation. Examples are \cite{Duary:2026izp,Sheikh-Jabbari:2026rcr,Bagchi:2026pjx,Sheikh-Jabbari:2026nez,Uvarov:2025whq}. 

The focus has been on studies in flat backgrounds. Here we shift it to $AdS$ backgrounds which may be of relevance, eg., for higher spin theories \cite {Fotopoulos:2006ci} and for limits in AdS/CFT \cite{Haggi-Mani:2000dxu,Sundborg:2000wp}. Related studies are 
\cite{Lindstrom:2026tua,Sheikh-Jabbari:2026rcr,Duary:2026izp,Duary:2026zar}.

Our construction relies on Dirac's embedding space, previously used to construct the null string \cite{Karlhede:1986wb} and the conformal string \cite{Gustafsson:1994kr}. Whereas the immediate generalisation of the conformal string to $AdS$ lacks local scaling invariance, it may be included via a St\"uckelberg field. This model is referred to as the $eAdS$ string.

In the reduced $d+1$ dimensional space one may alternatively consider the gauged ILST string in an $AdS$ background. The $eAdS$ string is crucially different from that. The local scaling transformations of the gauged ILST action do not commute with the $so(d,2)$ symmetry of the ambient $AdS$ space\footnote{The same feature makes the gauged ILST string incompatible with Poincar\'e invariance \cite{Lindstrom:2026tua}.}, contrary to the the local scaling transformations of the $eAdS$ formulation which commute with the $so(d,2)$ isometries. 

That there are two alternative gaugings is relevant for several reasons.  
The Carrol-Weyl symmetric action (gauged ILST) in an AdS background reduces to an ordinary tensionless string in one dimension less, and is not the tensionless limit of a tensile string in AdS.
The gauged ILST model consequently loses degrees of freedom compared to a tensionless string. Our new gauging does lead to a Lagrangian that has all the isometries of the $AdS$ space time, and  can be derived from a tensile string model; a tensile string in $AdS$ 
with an extra 
world sheet scalar field.\footnote{The extra scalar is charged and coupled to a  one-form which is closed on-shell.}

\section{The  $AdS_{d+1}$ null string embedded in $d+2$ dimensions}

The Dirac construction of conformal models proceeds via an ${d+2}$ dimensional Minkowski like embedding space with signature $-+++....+-$. The model is then formulated in this higher dimensional space together with an embedding condition. Solving this condition introduces the lower dimensional coordinates and defines the lower dimensional model. Here we take the $d+2$ model to be
\beq\label{Dact2}
S_{AdS}=\int d^2\xi \big(V^\al V^\be {\cal D}_{\al}  X^{M}{\cal D}_{\be} X^{N}\hat \eta_{MN}\big)+\Phi ( X^2+R^2e^\varphi)~.
\eeq
where $X^M~,M=0,1,...,d+1$ are embedding space, $\xi^\al = (\tau, \sigma)$ are world sheet coordinates, $\Phi$ is a Lagrange multiplier for the embedding constraint and $\varphi$ is a St\"uckelberg field introduced to ensure local scaling invariance. The local symmetry that leaves the action invariant is, infinitesimally,
\begin{align}
\label{GCons}
\de X^M=\lambda X^M,~~
\de V^\al=-\lambda V^\al,~~
\de W_\al=-\partial_\al\lambda,~~
\de\Phi=-2\lambda \Phi,
~~\de \varphi=2\lambda.
\end{align}
and the corresponding covariant derivative reads
\beq
 {\cal D}_{\al}  X^{M}=\partial_\al X^{M}+W_\al X^{M}
\eeq
with $W_\al$ the gauge potential.\\

Comment: There are two immediate extremes, $\varphi=0$ and $\varphi=-\infty$. The first turns the action into that of the ungauged $AdS_{d+1}$ string with embedding condition $X^2=-R^2$ breaking the local scaling invariance.  The second leads to the conformal string with embedding condition $X^2=0$ preserving local scaling invariance. In addition to this, changing the sign on the $R^2$ term will lead to a gauge invariant embedded $dS$ string.

\section{The $eAdS$ null string as a tensionless limit}
\label{DRed}

The action \re{Dact2} may be arrived at by taking the tensionless ($T\to 0$) limit of 
\beq
\label{TDact2}
S=-{\textstyle \frac 1 2}\int d^2\xi\Big(\,Te^{-\varphi}\sqrt{-h}\,h^{\alpha\beta}{\cal D}_\alpha X^M{\cal D}_\beta X^N\hat\eta_{MN}-2\Phi\big(X^2+R^2e^\varphi\big)\Big),
\eeq
with $h_{\al\be}$  the world sheet metric. This is a tensile string in the embedding space coupled to the St\"uckelberg field $\varphi$. Its local scaling invariance is a subset of \re{GCons}:
\begin{align}
\label{TGCons}
\de X^M=\lambda X^M,~~
\de W_\al=-\partial_\al\lambda,~~
\de\Phi=-2\lambda \Phi,
~~\de \varphi=2\lambda.
\end{align}
The relation between the metric $h$ and vector density $V$ is derived for the Nambu-Goto string in \cite{Lindstrom:1990qb} and we adapt it here to read
\beq
\label{hinv}
h^{\al\be}=-4\kappa(V^\al V^\be- \kappa T^2e^{2\varphi}\delta^\al_\sigma \delta^\be_\sigma)
\eeq
where $\kappa$ is related to the determinant of the metric by
\beq
\label{hdet}
\sqrt{-h}=\frac {e^\varphi}{2\kappa T}.
\eeq
Inserting \re{hinv} and \re{hdet}  into \re{TDact2} yields the tensionless $eAdS$ string \re{Dact2} when $T\to 0$.\\

There are several further aspects of the $eAdS$ string we will discuss. The three main ones are: 
~~i) Reduction to lower dimension,~~ ii) The commutation of local scaling and embedding space transformations, and ~~iii)
Hamiltonian analysis. We next treat the reduction.

\section{Reduction of $eAdS_{d+2}$ strings to $d+1$ dimensions }

There are two main ingredients to the reduction: the solution of the embedding constraint and the elimination of the gauge field $W$. The $\Phi$ equation is the embedding constraint $X^2+R^2e^\varphi=0$. This is solved by the choice 
\beq
\label{embsln}
X^M=(X^m,X^{d+1})=e^{\frac \varphi 2}(x^m,\sqrt{x^2+R^2})~.
\eeq
Note that this implies that $x^m$ are inert under the local scaling transformations 
\re{TGCons}.
We then have the choice of implementing these coordinates before or after eliminating $W$. For completeness, we present both routes. 
For the Hamiltonian analysis later, we need to keep $W$ in the action. So, first, using  only the solution \re{embsln}, we derive
\beq
\label{AdGILST}
S=\int d^2\xi e^{\varphi}V^\al V^\be\big( {\partial}_{\al} x^m{\partial}_{\be}  x^n
g_{mn}-R^2\Omega_\al\Omega_\be\big),
\eeq
with 
\beq
\Omega_\al=\textstyle \frac 1 2\partial_\al\varphi +W_\al
\eeq
 and induced $AdS_{d+1}$ metric 
\beq
g_{mn}=\eta_{mn}-\frac{x_mx_n}{x^2+R^2}~.
\eeq

Second, eliminating the gauge field first instead takes the us to the space of orbits, i.e. the quotient space. The $W$ equations are 
\beq
0=X^MV{\cal D}X^N\hat\eta_{MN}={\textstyle \frac 1 2}V\partial X^2+VW X^2.
\eeq
Using the embedding condition this implies that the vanishing of the scaling invariant combination
\beq
V\Omega:={\textstyle \frac 1 2}V\partial \varphi+VW
\eeq
In deriving this we did not use the explicit solution \re{embsln}, only the condition itself. So the quotient is given by
\beq
S_{W'}=\int d^2\xi \big(V^\al V^\be {\cal D}_{\al}  X^{M}{\cal D}_{\be} X^{N}\hat \eta_{MN}\big)
\eeq
where $W=-{\textstyle \frac 1 2}\partial \varphi$. Inserting the solution \re{embsln} in this we finally find 
\beq
\label{eAdSreduced}
S_{W'}\to\int d^2\xi e^{\varphi}\big(V^\al V^\be {\partial}_{\al}  x^{m}{\partial}_{\be} x^{n}g_{mn}\big)
\eeq
This agrees with eliminating $W$ from \re{AdGILST}.

The tensile string \re{TDact2} reduces similarly to
 \beq
\label{TAdGILST}
S=-{\textstyle \frac 1 2}\int d^2\xi e^\varphi e^{-\varphi}T\sqrt{h}\,h^{\al\be}\big( {\partial}_{\al} x^m{\partial}_{\be}  x^n
g_{mn}-R^2\Omega_\al\Omega_\be\big)
\eeq
using the solution of the embedding condition.
The worldsheet metric $h$ given by \re{hinv} and \re{hdet} confirms that the tensionless limit is \re{AdGILST}, resulting in \re{eAdSreduced} after eliminating $W$.  

\section{Local scaling and embedding space transformations commute}
The symmetries of the $d+2$ dimensional embedding space are $SO(d,2)$ rotations with generators $\omega$ that act on the $d+2$ coordinates as $X^M=\omega^M_{~N}X^N$. Since the $\omega$ matrices are constant and $X^2$ invariant, it immediately follows that the action \re{Dact2} is invariant under these rotations. In particular, the embedding constraint $\varphi =ln\big(\frac{-X^2}{R^2}\big)$ is invariant and thus $\varphi$ is $so(d,2)$ invariant as are $V$ and $W$, since the action is invariant without them transforming. This action is thus invariant under both local scaling transformations and embedding space isometries. Since the $\varphi, V^\al$ and $W_\al$ do not transform under the $so(d.2)$, they commute. We now show that this is still true for the reduced theory.

The $X$ coordinates that solve the embedding conditions are 
\beq
X^M=(X^m,X^{d+1})=e^{\textstyle \frac \varphi 2}(x^m,\sqrt{x^2+R^2})=:e^{\textstyle \frac \varphi 2}Y^M~.
\eeq
From $\delta X^M=\omega^M_{~N}X^N$ and $\varphi$ a scalar, it follows that $\delta X^M=e^{\textstyle \frac \varphi 2}\delta Y^M=(\omega^M_{~N}Y^N)e^{\textstyle \frac \varphi 2}$. The transformations split into Lorentz transformations and translations as
\begin{align}\nn 
\label{atfs}
&\delta_L x^m=\omega^m_{~n}x^n~~~ \makebox{Lorentz}\\\nn
&\delta_T x^m=a^m\sqrt{x^2+R^2}~~~ \makebox{$d+1$ AdS translation}\\
&\Rightarrow \delta_T \sqrt{x^2+R^2}=a_mx^m, ~~~\makebox{consistency}
\end{align}
Using the transformations \re{atfs} it is straightforward to show that \re{AdGILST} is invariant with $\varphi, V$ and $W$ as spectators. Since it is also invariant under local scaling transformations
\beq
\delta V^\al=-\lambda V^\al~,~~~\delta \varphi = 2\lambda
\eeq
while $x^m, g_{mn}$ and $\Omega$ are invariant, 1.e., act as spectators,
we see that the local scaling transformations and ambient space isometries again commute.

 \section{Hamiltonian analysis}

\subsection{Constraint analysis of the $eAdS$ null string in $d+2$ dimensions}
 \label{Sconstraints}
 
 From \re{Dact2} it follows that the momenta conjugate to $ X^M, (M=0,1,...,{d+1})$  are
 \beq
 P_M=2V^\tau V\cdot {\cal D} X_M~.
 \eeq
 Solving this for $\dot{X}^M$ we find the canonical Hamiltonian to be
 \beq
 {\cal H}_C=P\cdot \dot{X}-{\cal L}=\frac {P^2}{4(V^\tau)^2}-\frac {V^\sigma}{V^\tau}P_M{\cal D}_\sigma X^M-\Phi (X^2+R^2e^\varphi)~~.
 \eeq
 In addition there are six primary constraints due to the absence of some momenta
 \beq
 P_{V^\tau}=0~, ~~P_{V^\sigma}=0~,~~P_{W_\tau}=0~, ~~P_{W_\sigma}=0~,~~P_\Phi=0~,~\makebox{and}~~P_\varphi=0~.
 \eeq
 Following the Dirac procedure we include these in the Hamiltonian using Lagrange multipliers
 \beq
 {\cal H}_T={\cal H}_C+\lambda^\al P_{V^\al}+\mu^\al  P_{W_\al}+\lambda^\Phi P_\Phi+\lambda^\varphi P_\varphi~.
 \eeq
 Stability, or Dirac consistency, now requires that $[{\cal H}_T,\bullet ]=0$ for the Poisson bracket of ${\cal H}_T$ with any of the primary constraints, represented by $\bullet$. This leads to the secondary constraints collected in the following table:
 
\begin{center}
\begin{tabular}{|c|c|c|}
\hline
Primary& Secondary& Interpretation\\
\hline
~&~&~\\
$~~P_{V^\tau}\approx 0~~$ & ~~~~~$\frac{1}{2(V^\tau)^3}P^2-\frac{V^\sigma}{(V^\tau)^2}
P\cdot {\cal D}_\sigma X\approx 0$ & Mass shell condition. ($P^2\approx 0$)\\
~&~&~\\
$P_{V^\sigma}\approx 0~~$&$\frac 1 {V^\tau}{P\cdot} {\cal D}_\sigma X\approx 0$&$\sigma $ reparametrization (${P\cdot} {\cal D}_\sigma X\approx 0$)~\\
~&~&~\\
 $P_{W_\tau}\approx 0~~$&$P\cdot X\approx 0$&Generator of scaling transformations\\
 ~&~&~\\
$P_{W_\sigma}\approx 0~~$&$V^\sigma P\cdot X\approx 0$&Redundant\\
~&~&~\\
$P_{\Phi}\approx 0~~$&$X^2+R^2e^\varphi \approx 0$&  $AdS$  embedding constraint\\
~&~&~\\
$P_{\varphi}\approx 0~~$&$\Phi R^2e^\varphi\approx 0$& $\Phi\approx 0$ \\
~&~&~\\
\hline
\end{tabular}
\end{center}
\label{default}

Note that due to the constraint $PX\approx 0$ in the third row, the constraint ${P\cdot} {\cal D}_\sigma X$ is equivalent to $\phi^L:=P\cdot X'\approx 0$, which is our $\sigma$  reparametrizations constraint. 

To proceed further we need to separate the constraints into first and second class. If we had $\varphi\to -\infty$, the gauge algebra would be generated by the four first class constraints $\phi^{-1}:=P^2, ~\phi^0:=P\cdot X, ~\phi^L:=P\cdot X'$ and $X^2$. Here $X^2$ is modified into  $\chi_1:=X^2+R^2e^\varphi$ which together with $\chi_2:=\Phi, P_\Phi$ and $P_\varphi$ form a quadruple of second class constraints. We would want the remaining constraints $\phi^{-1},~ \phi^0,~\phi^L$ to be first class and generate the gauge algebra. This is not quite the case since $P\cdot X'$ and $PX$ have non-vanishing brackets with the $\chi_1$. To remedy this we recall a technique from Dirac's formalism: we shift the first-class constraint by a linear combination of the second-class constraints so that its Poisson bracket with all second-class constraints vanishes weakly.

Our modified constraints can be written
\beq
{\cal C}_1=P^2~,~~ {\cal C}_2=PX'-\varphi 'P_\varphi~,~\makebox{and} ~~{\cal C}_{PX}=PX-2P_\varphi~,
\eeq
Together with the momenta $P_V, P_W$ there are seven first class constraints. 
The four second class constraints are
\beq
\chi_1:=X^2+R^2e^\varphi~,~ P_\varphi~,~ \Phi~ \makebox{and}~~P_\Phi.
\eeq
We can now count the physical phase space degrees of freedom. In addition to the $X^M$ we have the six fields $V^\alpha,W_\alpha,\Phi$ and $\varphi$. Seven first class constraint removes two degrees of freedom each and four second class constraint removes one degree of freedom each. Thus, we have $2\times (d+2)+2\times 6 - (2\times 7)-4=2\times (d-1)=2\times (d+1-2)$ physical degrees of freedom, as expected of a string in $AdS_{d+1}$.

The four second class constraints $\chi_1, P_\varphi, \Phi$ and $P_\Phi$ can be eliminated once we defined the Dirac brackets. From their Poisson brackets we find the matrix
\beq\label{Dirac}
C(\sigma,\sigma')=\left(\begin{array}{cc}
0&{R^2e^\varphi}\\
-R^2e^\varphi&0
\end{array}\right)_{\!\!\!\!(\chi_1,P_\varphi)}\!\!\!\!\!\!\!\!\oplus
\left(\begin{array}{cc}
0&1\\
-1&0
\end{array}\right)_{\!\!\!\!(\Phi,P_\Phi)}\!\!\!\!\!\!\!\!\delta(\sigma-\sigma')
\eeq with inverse
\beq\label{DiracInv}
C^{-1}
(\sigma,\sigma')=\left(\begin{array}{cc}
0&-\frac 1{R^2e^\varphi}\\
\frac 1{R^2e^\varphi}&0
\end{array}\right)\oplus
\left(\begin{array}{cc}
0&-1\\
1&0
\end{array}\right) \delta(\sigma-\sigma')
\eeq 
such that 
\beq
\int d\sigma'' C_{ik}(\sigma,\sigma'')C^{-1}_{kj}(\sigma'',\sigma')=\delta_{ij}\delta(\sigma-\sigma')~.
\eeq
The Dirac bracket between $A$ and $B$ is
\beq
[A,B]_D=[A,B]-\int d\sigma_1d\sigma_2[A,c_i(\sigma_1)]C^{-1}_{ij}(\sigma_1,\sigma_2)[c_j(\sigma_2),B]
\eeq
where $c_i$ runs over the four second class constraints. Explicitly,
\beq
[A,B]_D=[A,B]+\int d\sigma\Big[\tfrac1{R^2e^\varphi}\big([A,\chi_1][P_\varphi,B]-[A,P_\varphi][\chi_1,B]\big)+[A,\Phi][P_\Phi,B]-[A,P_\Phi][\Phi,B\}\Big]~.
\eeq
Although the Dirac bracket permits calculations in the larger phase space of $X, P, \varphi,P_\varphi,\Phi,P_\Phi$ we know from the counting above that $X,P$ suffice, supplemented by three first class constraints. 

\subsection{Constraint analysis of the $eAdS$ string in $d+1$ dimensions.} 

We first derive the constraints that follow from the model \re{AdGILST}. 

 The momenta derived from the Lagrangian in \re{AdGILST} are
 \beq
 p_m=2e^\varphi V^\tau (V\partial x^n)g_{nm}~,~~p_\varphi=-e^\varphi R^2V^\tau (V\Omega)~.
 \eeq
The canonical Hamiltonian reads
\beq
H_C=\frac {e^{-\varphi}}{(V^\tau)^2}\Big(\frac 1 4 p^2-\frac 1 {R^2}p^2_\varphi\Big)-\frac {V^\sigma}{V^\tau}\Big(p_mx'^m+2p_\varphi\Omega_\sigma\Big)-2p_\varphi W_\tau
\eeq
The four momenta $p_V$ and $p_W$ are absent from $H_C$ which leads to the total Hamiltonian being
\beq
H_T=H_C+\lambda_Vp_V+\lambda_Wp_W~.
\eeq
Consistent time development requires the brackets $\{H_T,p_V\}$ and $\{H_T,p_W\}$  to vanish. This leads to the secondary constraints\footnote{We use the freedom to consider linear combinations to find these shifted forms.}
\beq
\mathcal C_1:=p^2-\frac 4 {R^2} p_\varphi^2\approx 0~,~~~\mathcal C_2:=px'+p_\varphi\varphi'\approx 0~,~~~\mathcal C_\varphi:=p_\varphi\approx 0.
\eeq
They obey
\begin{align}
\label{Ficonst}\nn
&\{\mathcal C_1(\sigma),\mathcal C_1(\sigma')\}=0\\\nn
&\{\mathcal C_2(\sigma),\mathcal C_2(\sigma')\}=(\mathcal C_2(\sigma)+\mathcal C_2(\sigma'))\delta'(\sigma-\sigma'),\\\nn
&\{\mathcal C_\varphi(\sigma),\mathcal C_2(\sigma')\}=\mathcal C_\varphi(\sigma')\delta'(\sigma-\sigma')\\\nn
&\{\mathcal C_\varphi(\sigma),\mathcal C_\varphi(\sigma')\}=0~,\\\nn
&\{\mathcal C_1(\sigma),\mathcal C_2(\sigma')\}=(\mathcal C_1(\sigma)+\mathcal C_1(\sigma'))\delta'(\sigma-\sigma'),
\\
&\{\mathcal C_\varphi(\sigma),\mathcal C_1(\sigma')\}=0
\end{align}
We see that the constraints are first class and, as will be discussed elsewhere,  that the algebra is formally the same as that of the gauged ILST string in an  $AdS$  background \cite{LindstromSundborgInPrep}. 
The whole model is invariant under scaling transformations as seen using \re{GCons}. We note  that the bracket of $\mathcal C_\varphi$ and $\mathcal C_1$ vanishes, signaling  that the commutator of the Fourier modes $[M_m,S_n]$ vanishes, corresponding to a  contraction of the Carroll-Weyl algebra, analogous to the result \cite{Duary:2026zar}  for the gauged ILST string.

Here, the absence of second class constraints leads to $2(d+1)+2+2\times 4-2\times (3+4)=2(d+1-2)$ phase space degrees of freedom, consistent with the result in section \ref{Sconstraints}, obtained before the reduction to $d+1$ dimensions. Thus, the reduction is just a different description of the same physics. In contrast to the gauged ILST string in $d+1$
\beq\label{wred}
S\to \int d\xi^2\left( V\cdot{\cal D}\,x^m g_{mn}(x)V\cdot{\cal D}\,x^n \right) ~,
\eeq 
the present model has a worldsheet scalar $\varphi$ which restores degrees of freedom removed by the extra local scaling symmetry. Fixing the scaling gauge removes $C_\varphi$ but also the $\varphi$ degrees of freedom, leaving a standard null string in $AdS_{d+1}$.

\section{Conclusions/Summary}
The construction of the $eAdS$ string may be adapted to cover the un-gauged $AdS$ ($\varphi =0$), the conformal string ($\varphi =-\infty$) or the gauged $dS$ string ($R^2\to -R^2$).  Here we have analysed the $AdS$ case.

The presented $eAdS$ formulation is locally scaling symmetric, or ``Carroll-Weyl" symmetric, a symmetry lacking in the ILST formulation, which was derived from the tensionless limit of standard strings. One of our main findings, that the $eAdS$ null string is also a tensionless limit, shows the advantage of the new $eAdS$ formulation: It combines a proper local symmetry with a relationship to ordinary strings.  
In addition, the $eAdS$ formulation preserves all the isometries of AdS$_{d+1}$, since the ambient $SO(d,2)$ commutes with the local scaling symmetry. It propagates $d-1$ transverse modes, the same as an ordinary string in $AdS_{d+1}$. This differs from the gauged ILST string in $d+1$ dimensions \cite{Sheikh-Jabbari:2026vqh,Sheikh-Jabbari:2026tpf,Lindstrom:2026zno}, which removes one mode and leaves $d-2$, those of an ILST string in $d$-dimensional Minkowski space. The St\"uckelberg field $\varphi$ restores that mode, and fixing $\varphi=0$ gives the standard null string in AdS$_{d+1}$.

Another new feature we have emphasised, which sets it apart from the gauged ILST string, is that the $eAdS$ string is the tensionless limit of a particular tensile string, both in the $d+2$ dimensional embedding space and directly in 
$AdS_{d+1}$. 
We hope to return to this interesting feature elsewhere.
\clearpage

\Large

\centerline {\bf Appendix}
\normalsize
\bigskip

\appendix

\section{ $AdS_{d+1}$  and some of its coordinates}
\label{AdsGeom}
Anti deSitter space in $d+1$ dimensions can be realized as a homogeneous space, the coset space 
\beq
 AdS_{d+1}=\frac {SO(d,2)}{SO(d,1)}~.
\eeq{}
We will make use of an explicit construction adapted from the Dirac Conformal space space. So consider $\mathbb{R}^[d.2]$  with coordinates $X^M$, $M=0,1,...,d,d+1$ and metric signature $(-,+++...+,-)$. Imposing the embedding constraint 
\beq\label{emb}
X^2=\sum_{m=1}^{d}(X^m)^2-(X^0)^2-(X^{d+1})^2=-R^2
\eeq
reduces the dimension to $d+1$.

Solving \re{emb} can be made using different coordinates. The coordinates used in the present paper are sometimes called ``Beltrami coordinates''. We give them below and their relation to the perhaps more well known Poincar\'e coordinates.

\subsection{Beltrami coordinates.}

We  reduce from $d+2$ to $d+1$ by solving  the embedding condition for $X^{d+1}$:
\beq\label{weier}
X^{d+1} = \pm \sqrt{X^2+R^2}~, 
\eeq
where $X^2:=X^mX^n\eta_{mn}$, $m,n=0,1,...d$. The induced $d+1$ dimensional line element then reads
\beq\label{Winduced}
ds_{|}^{2}=dX^m\Big(\eta_{mn} -\frac{X_{m}X_{n}}{X^{2}+R^{2}}\Big)dX^n~ =: dX^m\hat\eta_{mn}dX^n~,
\eeq
which is recognizable as the metric on the space of orbits for the isometry, the bulk isometry that scales the vector \(X^{d+1} + X^0 \). This is an example of a reduction (quotient) under an isometry. 

The relation to the Poincar\'e coordinates is 
\begin{align}
&X^{d+1} + X^0 = \frac{R^2}{z}\\\nn
&X^{d+1} - X^0 = -z - \frac{\eta_{\mu\nu}x^\mu x^\nu}{z}\\
&X^\mu = \frac{R x^\mu}{z} ~,
\end{align}

where (\(\mu = 1, \dots, d\)).

\section{Equations for the reduced $eAdS$  string in $d+1$ dimensions}
 \label{EOM}

 We copy the new model \re{AdGILST} for convenience.
 \beq\label{2AdGILST2}
S=\int d^2\xi e^{\varphi}\big(V^\al V^\be {\partial}_{\al}  x^m{\partial}_{\be}  x^n
g_{mn}-R^2V^\al V^\be\Omega_\al\Omega_\be\big).
\eeq
It may be thought of as a string or sigma model with a degenerate world sheet metric, coupled to a dilaton. The $x^m$ equations are 
\beq
\partial_\beta\big(V^\alpha V^\beta\partial_\alpha x^p\big)+\Gamma^p_{mk}\,V^\alpha V^\beta\partial_\alpha x^m\partial_\beta x^k+V^\alpha V^\beta\partial_\alpha x^p\,\partial_\beta\varphi=0~,
\eeq
which we recognize as the harmonic map for a sigma model with a degenerate inverse metric $h^{\al\be}=V^\al V^\be$. ($\Gamma$ is the Levi-Civita connection for $g$). 

The $V^\al$  equations are
\beq
e^\varphi V^\be\big(\gamma_{\al\be}-R^2\Omega_\al\Omega_\be\big)=:V^\be T_{\al\be},
\eeq
where $\gamma_{\al\be}:={\partial}_{\al}g_{mn}  x^m{\partial}_{\be}  x^n
$ is the usual induced metric and $T_{\al\be}$ is the energy momentum tensor (cf. \cite{Isberg:1993av}).

The $W_\al$ equations are
\beq
-2e^\varphi  V^\al V\cdot\Omega=0
\eeq

The $\varphi$ equation is
\beq
{\cal L}+R^2\partial_\al\big(e^\varphi V^\al V\cdot\Omega\big)=0~,
\eeq
where ${\cal L}=V^\al V^\be \big(\gamma_{\al\be}-R^2\Omega_\al\Omega_\be\big)=V^\al V^\be T_{\al\be}$
Combining these equations, we find that, on shell,  $\gamma$ is degenerate (the null condition), ${\cal L}=0$ and $V^\al W_\al=-\textstyle\frac 12 V^\al\partial_\al \varphi$
~. When $h$ is non-degenerate, this reverts to  familiar results for a tensile string. Hence, our model is indeed the tensionless limit of a  tensile string in curved (conformal $AdS$) background, in agreement with the discussion in subsection \ref{DRed}.

\setcounter{page}{1}

\bibliographystyle{JHEP}
\bibliography{references}

\providecommand{\href}[2]{#2}\begingroup\raggedright\begin{thebibliography}{10}

\bibitem{Schild:1976vq}
A.~Schild, \emph{{Classical Null Strings}}, \href{https://doi.org/10.1103/PhysRevD.16.1722}{\emph{Phys. Rev. D} {\bfseries 16} (1977) 1722}.

\bibitem{Karlhede:1986wb}
A.~Karlhede and U.~Lindstr{\"o}m, \emph{{The Classical Bosonic String in the Zero Tension Limit}}, \href{https://doi.org/10.1088/0264-9381/3/4/002}{\emph{Class. Quant. Grav.} {\bfseries 3} (1986) L73}.

\bibitem{Isberg:1993av}
J.~Isberg, U.~Lindstr{\"o}m, B.~Sundborg and G.~Theodoridis, \emph{{Classical and quantized tensionless strings}}, \href{https://doi.org/10.1016/0550-3213(94)90056-6}{\emph{Nucl. Phys. B} {\bfseries 411} (1994) 122} [\href{https://arxiv.org/abs/hep-th/9307108}{{\ttfamily hep-th/9307108}}].

\bibitem{Duary:2026izp}
S.~Duary and S.~Maji, \emph{{Null string holography in AdS from local BRST descent}},  \href{https://arxiv.org/abs/2610.10540}{{\ttfamily 2610.10540}}.

\bibitem{Sheikh-Jabbari:2026rcr}
M.M.~Sheikh-Jabbari and H.~Yavartanoo, \emph{{Null String Holography, Null Strings Probe Projective Boundary of Their Target Spaces}},  \href{https://arxiv.org/abs/2609.10432}{{\ttfamily 2609.10432}}.

\bibitem{Bagchi:2026pjx}
A.~Bagchi, S.~Grover, S.~Rajesh~Iyer, A.~Saha and S.~Stieberger, \emph{{Double Copy from the Flipped Null String}},  \href{https://arxiv.org/abs/2609.07846}{{\ttfamily 2609.07846}}.

\bibitem{Sheikh-Jabbari:2026nez}
M.M.~Sheikh-Jabbari and H.~Yavartanoo, \emph{{Carrollian Worldsheets on Carrollian Targets, When does ILST describe a null string theory?}},  \href{https://arxiv.org/abs/2610.03495}{{\ttfamily 2610.03495}}.

\bibitem{Uvarov:2025whq}
D.V.~Uvarov, \emph{{On Integrable Structure of Null String in (Anti-)de Sitter Space}}, \href{https://doi.org/10.3842/SIGMA.2026.005}{\emph{SIGMA} {\bfseries 22} (2026) 005} [\href{https://arxiv.org/abs/2508.04158}{{\ttfamily 2508.04158}}].

\bibitem{Fotopoulos:2006ci}
A.~Fotopoulos, K.L.~Panigrahi and M.~Tsulaia, \emph{{Lagrangian formulation of higher spin theories on AdS space}}, \href{https://doi.org/10.1103/PhysRevD.74.085029}{\emph{Phys. Rev. D} {\bfseries 74} (2006) 085029} [\href{https://arxiv.org/abs/hep-th/0607248}{{\ttfamily hep-th/0607248}}].

\bibitem{Haggi-Mani:2000dxu}
P.~Haggi-Mani and B.~Sundborg, \emph{{Free large N supersymmetric Yang-Mills theory as a string theory}}, \href{https://doi.org/10.1088/1126-6708/2000/04/031}{\emph{JHEP} {\bfseries 04} (2000) 031} [\href{https://arxiv.org/abs/hep-th/0002189}{{\ttfamily hep-th/0002189}}].

\bibitem{Sundborg:2000wp}
B.~Sundborg, \emph{{Stringy gravity, interacting tensionless strings and massless higher spins}}, \href{https://doi.org/10.1016/S0920-5632(01)01545-6}{\emph{Nucl. Phys. B Proc. Suppl.} {\bfseries 102} (2001) 113} [\href{https://arxiv.org/abs/hep-th/0103247}{{\ttfamily hep-th/0103247}}].

\bibitem{Lindstrom:2026tua}
U.~Lindstr{\"o}m and B.~Sundborg, \emph{{(Anti-)De Sitter null strings and Carroll-Weyl symmetry}},  \href{https://arxiv.org/abs/2609.07765}{{\ttfamily 2609.07765}}.

\bibitem{Duary:2026zar}
S.~Duary and S.~Maji, \emph{{Carroll-Weyl gauged null strings in AdS}},  \href{https://arxiv.org/abs/2610.10717}{{\ttfamily 2610.10717}}.

\bibitem{Gustafsson:1994kr}
H.~Gustafsson, U.~Lindstr{\"o}m, P.~Saltsidis, B.~Sundborg and R.~von Unge, \emph{{Hamiltonian BRST quantization of the conformal string}}, \href{https://doi.org/10.1016/0550-3213(95)00051-S}{\emph{Nucl. Phys. B} {\bfseries 440} (1995) 495} [\href{https://arxiv.org/abs/hep-th/9410143}{{\ttfamily hep-th/9410143}}].

\bibitem{Lindstrom:1990qb}
U.~Lindstr{\"o}m, B.~Sundborg and G.~Theodoridis, \emph{{The Zero tension limit of the superstring}}, \href{https://doi.org/10.1016/0370-2693(91)91726-C}{\emph{Phys. Lett. B} {\bfseries 253} (1991) 319}.

\bibitem{LindstromSundborgInPrep}
{U. Lindstr{\"o}m and B. Sundborg}, ``In preparation.''.

\bibitem{Sheikh-Jabbari:2026vqh}
M.M.~Sheikh-Jabbari and H.~Yavartanoo, \emph{{Null Strings Gauged and Reloaded, I: Null Strings Have Carroll-Weyl Gauge Symmetry}},  \href{https://arxiv.org/abs/2605.25817}{{\ttfamily 2605.25817}}.

\bibitem{Sheikh-Jabbari:2026tpf}
M.M.~Sheikh-Jabbari and H.~Yavartanoo, \emph{{Null Strings Gauged and Reloaded, II: Consistent Classical Treatment of the Null Strings}},  \href{https://arxiv.org/abs/2605.26822}{{\ttfamily 2605.26822}}.

\bibitem{Lindstrom:2026zno}
U.~Lindstr{\"o}m, \emph{{The conformal null string in $d+2$ and $d$ dimensions}},  \href{https://arxiv.org/abs/2606.22498}{{\ttfamily 2606.22498}}.

\end{thebibliography}\endgroup

\end{document}